\documentclass[letterpaper]{article} 
\usepackage{aaai2026}  
\usepackage{times}  
\usepackage{helvet}  
\usepackage{courier}  
\usepackage[hyphens]{url}  
\usepackage{graphicx} 
\usepackage{natbib}  
\usepackage{caption} 
\usepackage{algorithm}
\usepackage{algorithmic}

\usepackage{newfloat}
\usepackage{listings}
\DeclareCaptionStyle{ruled}{labelfont=normalfont,labelsep=colon,strut=off} 
\floatstyle{ruled}
\newfloat{listing}{tb}{lst}{}
\floatname{listing}{Listing}
\usepackage{xcolor}

\usepackage{amsmath}
\usepackage{siunitx}
\usepackage{tabularx}
\usepackage{array} 
\usepackage{threeparttable}
\usepackage{multirow}

\title{Quantifying Political Partisanship for Cross-Platform Analyses}
\author {
    Mia Ameen,
    Christopher G. Healey
}
\affiliations{
    North Carolina State University, Raleigh, NC, USA\\
    fameen@ncsu.edu, healey@ncsu.edu
}

\begin{document}

\maketitle

\begin{abstract}
Research on political polarization on social media depends on the ability to reliably measure partisanship in user-generated content. However, existing approaches are typically tailored to platform-specific properties, such as structural affordances or linguistic conventions, which hurts generalizability across platforms. This limitation is increasingly consequential as the social media ecosystem fragments and fringe, alt-tech platforms emerge alongside mainstream ones. We propose a text-based, platform-portable methodology for measuring political partisanship in social media posts, anchored by an external news-credibility signal. Posts are embedded using a transformer-based sentence encoder and clustered into topic groups, which are labeled using the aggregated AllSides media bias scores of cited news outlets. A partisanship axis is then constructed in the embedding space as the difference between centroids of oppositely labeled clusters, and individual posts are scored by projection onto this axis. We apply the method to a corpus of approximately 1.3 million posts collected from Bluesky and Truth Social during the six months preceding the 2024 U.S. presidential election, providing the first cross-platform comparison of partisanship distributions on these two ideologically asymmetric platforms. The resulting partisanship scores correlate significantly with held-out AllSides media bias scores both in-distribution and out-of-distribution on an independent Twitter corpus, and recover within-platform partisan dynamics that platform identity alone cannot explain.
\end{abstract}


\section{Introduction}

Whether political polarization is rising in democratic societies remains contested. Some accounts argue that the perception of widespread polarization in the United States is amplified by influential political figures and news media \cite{fiorina2004culture}, while others find evidence of a substantive increase at the general public level \cite{han_disappearing_2011, iyengar_affect_2012}. A growing perspective on this debate implicates social media in facilitating political discourse and influencing public communication~\cite{kubin_role_2021, lu_social_2025, overgaard_different_2025}.

Empirically characterizing this relationship requires solving a prior measurement problem. Political polarization is a property of a distribution---the separation between political groups along some dimension---and most claims about it assume that we can reliably assign \emph{partisanship} to individuals or content. Without a partisanship signal, questions about polarization, cross-cutting exposure \cite{bail_exposure_2018}, ``echo chambers" \cite{cinelli_echo_2021}, or platform-motivated radicalization \cite{schulze_2022_radicalization} cannot be investigated rigorously. These are not abstract concerns: partisanship shapes attitudes, for example, with Republicans and Democrats in the United States viewing each other's policies as threats to national well-being \cite{pew_political_2014}; it shapes behavior, influencing which news users share regardless of its accuracy \cite{pennycook_2021_attention} and elevating toxic non-political exchanges in partisan discussions \cite{mamakos_social_2023}. Finally, it shapes real-world outcomes, for example, when partisan associations were implicated in regional variations in U.S. COVID-19 mortality \cite{gollwitzer_2020_partisan}.

A substantial body of literature has proposed methods for inferring partisanship from social media data, including content-based classifiers, network- and community-based inference, and community-embedding based approaches. We review these techniques in detail in the next section. Despite their methodological diversity, existing methods share an important limitation: most methods are tailored to platform-specific properties in ways that limit cross-platform comparisons---whether by relying on structural affordances or on platform-specific linguistic conventions. As a result, partisanship scores for one platform are not always comparable to those for another.

This limitation has become more consequential as the social media ecosystem has fragmented. While most prior work has focused on established mainstream platforms such as X (formerly Twitter)~\cite{conover_political_2021}, Facebook~\cite{del_vicario_echo_2016}, and Reddit \cite{morini_toward_2021}, a parallel ecosystem of \emph{alt-tech} platforms---including Gab, BitChute, Parler, Bluesky, and Truth Social---has emerged in response to perceived ideological biases and moderation practices on mainstream platforms, explicitly positioning themselves as a space for alternative political discourse and minimal content moderation. Although these platforms host smaller user bases, evidence suggests both react to and influence mainstream political communication~\cite{dehghan_politicization_2022, dowling_2025_alt}. Truth Social, in particular, has played a notable role in contemporary U.S. political discourse, serving as the primary outlet for a primary presidential candidate during the 2024 election cycle~\cite{zhang_trump_2025}. Bluesky, by contrast, has emerged as a destination for users departing mainstream platforms due to changes in ownership and moderation, with what is assumed to be a primarily left-leaning user base \cite{failla_im_2024}. This asymmetry---a right-leaning and a left-leaning alt-tech platform---makes the pair particularly attractive for cross-platform partisanship analysis. Yet political discourse on Truth Social remains relatively underexplored in the literature.

We argue that progress on these questions requires a measurement of political partisanship that is \emph{portable}: comparable in interpretation across text from different platforms, without relying on platform-specific features. In this work, we propose such a method and apply it to posts on Bluesky and Truth Social during the May 1 to October 31, 2024 period leading to the 2024 U.S. presidential election.

Our approach proceeds in four stages. First, we collect posts from both platforms over the study period. Second, we represent all posts using a semantic embedding based on a transformer-based sentence encoder to semantically cluster them into topic groups. Third, we identify posts that cite news sources with established partisan leanings using the AllSides Media Bias ratings tool\footnote{\url{https://www.allsides.com/media-bias/ratings}}. We aggregate these citations at the cluster level to assign a partisanship label to each cluster. Finally, following \citet{waller_quantifying_2021}, we define a \emph{social dimension} as an axis in an embedding space spanned by clusters with semantically opposing labels. This allows us to project all posts onto the axis to obtain partisanship scores. This pipeline operates on posts' text alone---without relying on follower graphs, community structure, or hashtag conventions---and is guided by consistent external bias ratings. This ensures the same procedure can be applied to any text-based platform, with the resulting scores sharing a common interpretive frame across platforms.

Our social media analysis includes the following novel contributions:
\begin{itemize}
    \item A text-based partisanship measurement method, informed by external media bias ratings, that is validated to produce comparable scores across three platforms (Bluesky, Truth Social, and X).
    \item The first cross-platform comparative analysis of political discourse on Bluesky and Truth Social during the 2024 U.S. presidential election.
    \item Empirical evidence that our approach captures within-platform partisan dynamics rather than merely recovering platform identity, as demonstrated through temporal patterns in candidate-mentioning posts during the election.
\end{itemize}

\begin{figure*}[t]
    \centering
    \includegraphics[width=1\textwidth]{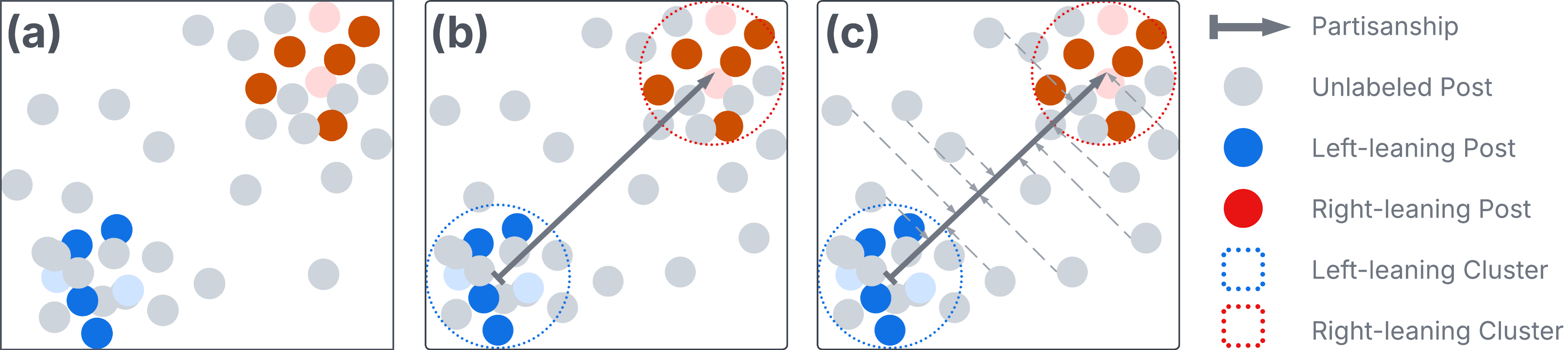}
    \caption{Overview of the social dimension construction pipeline. \textbf{(a)} Post embeddings prior to clustering, where blue and red points denote posts citing left- and right-leaning news media, respectively. \textbf{(b)} Clusters formed from these embeddings, labeled using the aggregated media bias scores of cited sources within each cluster. \textbf{(c)} Construction of the partisanship dimension as the vector difference between the centroids of two politically opposed clusters; individual partisanship scores for all posts are then computed by projecting each post onto this dimension.}
    \label{fig:method}
\end{figure*}

\section{Literature Review}

\subsection{Measuring Political Partisanship}
\label{sec:related}
The most direct approach to assessing an individual's political partisanship is to simply ask, typically via surveys such as a seven-point partisanship Likert scale~\cite{lee_social_2022}, or using established instruments such as the American National Election Studies, as \citet{ojer_charting_2025} do. Surveys yield high-quality individual labels but are expensive to administer and risk non-response and self-selection biases.


One workaround is to infer partisanship directly from post text. Dictionary methods rely on expert-curated keyword lists indicative of partisan intent, computing the proportion of keywords in a post to score its political leaning~\cite{hart_politicization_2020, simchon_troll_2022}. Hashtag-based methods exploit the convention that partisan users adopt distinctive tags~\cite{conover_partisan_2012}, while supervised classifiers train on labeled partisan corpora to learn lexical and syntactic markers of leaning~\cite{belcastro_learning_2020}. All three share two weaknesses: they require ongoing curation as platform-specific slang evolves, and they generalize poorly across platforms with different linguistic norms.

Another body of work leverages news source citations as a signal for partisanship, rather than platform-native conventions. Users' news-sharing behavior, with news outlets rated by tools such as Media Bias/Fact Check (MBFC), serves as a reliable proxy for political leaning \cite{weld_political_2021}. This is methodologically attractive because the signal lives outside the platform and can in principle be applied wherever users share news links.

A shared limitation of these text-based methods, however, is that they can only be applied to posts containing the signal of interest---a user who shares no news links or uses no partisan hashtags yields no estimate at all. Network-based methods address this gap by propagating partisanship labels via the relational structure of a social graph, reaching users who would otherwise go unscored. Such methods construct retweet, mention, follow, or like graphs and either apply community-detection algorithms such as Louvain or Leiden to recover politically homogeneous groupings, or propagate labels outward from a small set of nodes with known partisanship~\cite{rao_political_2021}. \citet{cinelli_echo_2021} and \citet{bassolas_multifaceted_2025} both use media bias ratings to directly assign partisan labels to citing nodes before inferring the positions of proximal nodes in the graph.  \citet{morales_measuring_2015}, in a related approach, use the accounts of politicians and journalists with publicly established political positions as the starting points for label propagation instead. Some methods exploit the sub-community structure of certain platforms: Reddit's subreddit system, in which communities such as \texttt{r/Republican} act as ready-made partisan bins, has supported a substantial body of polarization research~\cite{marchal_be_2022}. All of these approaches, however, require relational data to construct a social graph, which may not be consistently available across platforms. Where no such structure exists, topic-modeling approaches such as Structural Topic Modeling recover latent themes that are then manually labeled for political leaning and aggregated into post-level partisanship scores~\cite{yarchi_political_2021}, though this introduces a labor-intensive annotation step.

Our approach borrows from the embedding-based work of \citet{waller_quantifying_2021}, who learn community embeddings~\cite{martin2017community2vec} from Reddit's subreddit structure: every instance of a user commenting in a subreddit becomes a training example, with the subreddit playing the role of the ``word'' and the user playing the role of the ``context'' in a word2vec-style model, so that subreddits with similar user bases end up close together in the resulting space. To define a partisan ``social dimension,'' they begin with a hand-picked seed pair of ideologically opposed subreddits (\texttt{r/democrats} and \texttt{r/Conservative}), algorithmically augment it with nine additional pairs, and take the mean of the vector differences as the dimension axis; all communities are then scored by projection onto this axis. Notably, the pipeline makes no use of post text.

We adopt the social dimension construction from this line of work but depart from it in a crucial respect. Waller and Anderson rely on Reddit's subreddit structure for both halves of the pipeline: user--subreddit co-occurrences supply the embedding, and curated pairs of opposed subreddits supply the seed pairs. We construct both from post text alone---the embedding from a transformer-based sentence encoder, and the seed pairs from clusters whose aggregate cited media bias ratings place them at opposite ends of the AllSides spectrum. This shift---from platform-specific structural signals to platform-independent text---is what makes the resulting partisanship scores comparable across platforms.


\subsection{Alternative Social Media Platforms}
Most research on social media political discourse has concentrated on mainstream platformsm \cite{kubin_role_2021}, leaving a smaller but growing literature on the alt-tech ecosystem, consisting of platforms like Gab, BitChute, and Parler. \citet{lima2018inside} find Gab to be dominated by far-right discourse with strong echo-chamber characteristics. \citet{erickson2025content} document volumes of hate speech on BitChute, and \citet{baines2021scamdemic} trace the persistence of anti-vaccination conspiracies on Parler. Across these studies, alt-tech platforms emerge as venues where contested or harmful discourse propagates at scale, in the relative absence of the moderation pressures present on their mainstream counterparts.

Bluesky has also received research attention, in part because its open API and decentralized architecture make data collection feasible~\cite{balduf_looking_2024, failla_im_2024}. Existing evidence suggests its user base skews left-leaning: in a human-annotated validation set reported by \citet{rostami_politisky24_2025}, 87.6\% of users expressed opposition to Donald Trump and only 8.8\% expressed support, while 49.2\% expressed support for Kamala Harris and 11.2\% expressed opposition. This asymmetry is consistent with broader characterizations of Bluesky as politically homogeneous relative to mainstream platforms~\cite{salloum_politics_2025}: internal disagreements exist, but the platform as a whole occupies a narrower ideological range.
Truth Social, however, remains relatively understudied. Most prior work has focused on data collection rather than discourse analysis. \citet{gerard_truth_2023} introduced the first publicly available Truth Social dataset; a subsequent release extended coverage through 2024 \cite{TS24}; and \citet{ameen2026truthstanceannotateddatasetconversations} further enriched the corpus with comment data and conversational context. We build on this last release in our analysis. The small body of substantive findings on Truth Social is nonetheless consistent. \citet{shah_can_2024} finds that posts containing Wikipedia links receive much lower engagement than posts without them, implying that neutral-toned conversations attract little interest on the platform. \citet{casanova_beyond_2026} apply an LLM-based pipeline to identify election-related rumors and find that approximately one in 150 posts qualifies, suggesting that misinformation circulates at a high rate. \citet{hughes_echo_2025} report high modularity in the platform's interaction structure, indicating that its communities remain ideologically insular. Together, these findings establish Truth Social as an informationally consequential platform, yet its broader political discourse---and its relationship to other alt-tech platforms---remains largely unexamined.

\section{Method}

\subsection{Data}
This study draws on both existing datasets and newly collected data for Truth Social and Bluesky. We study content posted from May 2024 to October 2024, the six months preceding the 2024 U.S. presidential election. This was a time when political discourse was at its peak. Both datasets include post-level metadata---text, author identifiers, time of creation, and engagement statistics (likes, replies, and repost counts)---as well as author-level metadata, including usernames and follower/following counts.

\subsubsection{Truth Social.}
We use the dataset released by \citet{ameen2026truthstanceannotateddatasetconversations}, which builds on a publicly available Truth Social dataset of \citet{TS24}. This base dataset contains approximately 1.2M posts and comments collected between February and November 2024. It is licensed for non-commercial reuse with attribution\footnote{\url{https://creativecommons.org/licenses/by-nc/4.0/}}, focusing on political discourse and compiled using a hybrid collection strategy. Posts containing daily trending political hashtags (e.g., \textit{\#trump2024}) were scraped using a custom tool built on the Stanford Internet Observatory's open-source framework ``Truthbrush"\footnote{\url{https://github.com/stanfordio/truthbrush}}, while a predefined set of politically relevant keywords was continuously monitored and collected in the same manner. Additionally, comments were collected for a subset of these posts by \citet{ameen2026truthstanceannotateddatasetconversations}, using the same framework.

\subsubsection{Bluesky.}
Bluesky data were collected using the official AT Protocol Python SDK (\texttt{atproto}), querying the \texttt{app.bsky.feed.searchPosts} endpoint. We apply the same political keywords used for Truth Social as search filters for Bluesky from May 2024 through October 2024. We provide these keywords in the Appendix. 

\subsubsection{Preprocessing.}
Raw posts from both platforms were preprocessed using the following steps.
\begin{enumerate}
\item Posts with missing or empty content, and posts timestamped with a creation date falling outside of the target time window period (May–October 2024) were removed.
\item Posts with identical (\textit{author, text}) pairs were deduplicated, as these indicate automated or spam-like behavior.
\item Reposts (``retruths'' on Truth Social) were excluded, as they do not represent original user-generated content.
\item Posts were filtered for substantive textual content: a post was considered substantive if, after stripping hashtags, @mentions, emojis, and URLs, at least three tokens remained. Non-substantive posts were removed prior to analysis.
\item The remaining posts were cleaned by removing HTML markup, encoding artifacts, hashtags, @mentions, and emojis, with any remaining whitespace collapsed to produce the final cleaned text.
\end{enumerate}

\begin{figure}[h]
    \centering
    \includegraphics[width=0.47\textwidth]{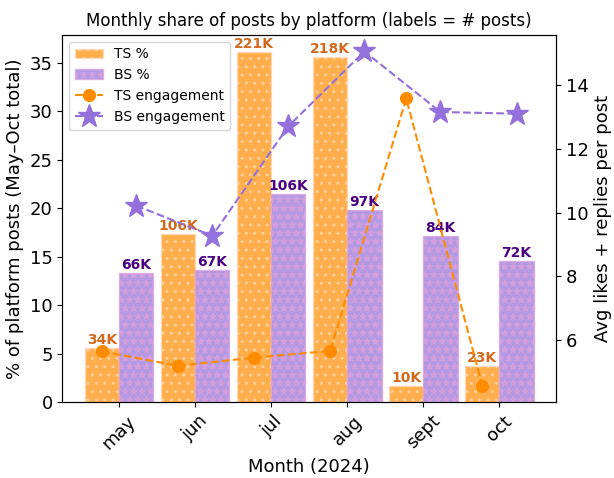}
    \caption{Monthly share of total posts and average engagement per post (likes + replies) for Truth Social (TS) and Bluesky (BS), May--October 2024. Bars indicate each platform's percentage of its own total posts within the observation window; dashed lines indicate average engagement on the secondary axis.}
    \label{fig:posts_per_month}
\end{figure}

Figure~\ref{fig:posts_per_month} shows the monthly distribution of posts, after pre-processing, across both platforms over the collection period. Both Truth Social and Bluesky show similar trends---activity is heavily concentrated in July and August, likely reflecting the heightened political discourse surrounding the Republican National Convention, U.S. president Joe Biden's withdrawal from the race, and Vice President Kamala Harris's announcement of her candidacy. 

\subsection{Partisanship Score Estimation}
To measure political partisanship at the post level, we introduce an unsupervised labeling pipeline.

\subsubsection{Embedding and Semantic Deduplication.}
All cleaned posts are embedded using \texttt{qwen3-embedding:8b}~\cite{zhang_qwen3_2025}, producing 128-d representations via Matryoshka truncation~\cite{kusupati2022matryoshka}. The model is served locally on an Nvidia A100 GPU using \texttt{Ollama}.\footnote{\url{https://github.com/ollama/ollama}} The resulting embeddings are L2-normalized before dimensional reduction using Principal Component Analysis (PCA), to retain projected components that collectively explain 95\% of the variance. (104 components).

Social media datasets can contain large amounts of redundant content, such as reposts, near-identical messages, and spam. We apply semantic deduplication independently to each platform-month partition following the procedure of \citet{abbas_semdedup_2023}. Posts are clustered using MiniBatch $k$-means with $k = 15$, and within each cluster, posts are sorted by descending cosine similarity to the cluster centroid. Pairwise cosine similarities are computed within each cluster, and for any pair of posts exceeding a similarity threshold of $0.90$, the post closer to the centroid is removed as a near-duplicate. The deduplication process is robust to the choice of $k$ provided it remains small relative to the overall dataset size. Inspecting the distribution of nearest-neighbor cosine similarities before and after deduplication confirmed that high-similarity neighbors were successfully removed. Table~\ref{tab:descriptive} reports the percentage of near-duplicates identified per platform. Notably, Truth Social exhibits a substantially higher duplication rate than Bluesky, suggesting a greater prevalence of redundant content, coordinated posting, or spam.

\subsubsection{Cross-Platform Clustering.}
Following deduplication, posts from both platforms are combined, jointly reduced in dimensionality, and clustered. MiniBatch $k$-means is again applied, with an elbow of $k=36$ selected by searching over $k \in [1, 500]$ and using the \texttt{kneed} package \cite{satopaa_finding_2011} to identify the point of maximum curvature on the inertia elbow curve.

\begin{table}[t]
\centering
\caption{Statistics of the Truth Social (TS) and Bluesky (BS) datasets after pre-processing and deduplication.}
\label{tab:descriptive}
\begin{tabularx}{\columnwidth}{@{}l@{\hspace{8pt}}r@{\hspace{8pt}}r@{\hspace{8pt}}r@{\hspace{8pt}}r@{\hspace{8pt}}r@{}}
\hline
\textbf{} &
\textbf{\#Posts} &
\textbf{\#Users} &
\textbf{\%Dup.} &
\textbf{\#Cit.} &
\textbf{Bias} \\
\hline
BS & $644,029$ & $119,590$ & $2.50\%$ & $22,936$ & $-1.34 \pm 0.83$ \\
TS & $672,215$ & $86,321$  & $8.82\%$ & $33,472$ & $1.25 \pm 0.95$ \\
\hline
\end{tabularx}
{
\textit{\#Posts} and \textit{\#Users} reflect counts after cleaning;
\textit{\%Dup.} is the share of posts removed as near-duplicates.
\textit{\#Cit.} is the total number of cited sources in each dataset;
\textit{Bias} is the mean AllSides bias score of cited sources $\pm$ std.}
\end{table}

\subsubsection{Cluster Labeling using News Domains.}

To assign political leanings to clusters, we leverage the media bias scores of news domains cited within a cluster's posts. We extract URLs from all posts via simple regex matching, compile the most frequently cited domains across both platforms, and then retrieve their media bias scores (when available) from AllSides, a publicly available online resource that provides ideology ratings for a news source on a five-point scale from $-2$ (far-left) to $+2$ (far-right). Table~\ref{table:news_counts} presents the five most frequently cited news sources and their ideology scores, for Bluesky and Truth Social. In total, our media bias dataset covers $104$ news domains, of which $48$ are labeled left-leaning ($-2$ or $-1$), $10$ as neutral ($0$), and $46$ as right-leaning ($+1$ or $+2$). Each post citing one or more domains is assigned the mean bias score of all the domains it references.

We estimate a cluster's political leaning by applying Empirical Bayes shrinkage to the media bias scores of its citing posts. This addresses the instability of raw means in clusters with only a few citations. For each cluster $c$ with $n_c$ cited posts, the raw mean $\bar{x}_c$ is contracted toward a global prior mean $\mu$ as follows:
\begin{equation*}
    \hat{\theta}_c = B_c \, \bar{x}_c + (1 - B_c) \, \mu, \quad
    B_c = \frac{\hat{\tau}^2}{\hat{\tau}^2 + \hat{\sigma}^2_c / n_c}
\end{equation*}

The prior mean $\mu$ and prior variance $\hat{\tau}^2$ are estimated as the sample mean and variance of the bias scores of all cited posts across all clusters. The per-cluster sampling-noise term $\hat{\sigma}_c^2 / n_c$ is the standard error of the cluster mean, where $\hat{\sigma}_c^2$ is the variance of bias scores across posts cited within cluster $c$. Clusters with many citations and consistent bias scores have $B_c \to 1$ and retain most of their observed mean, whereas clusters with limited citations or high bias score variation have $B_c \to 0$ and are pulled toward the global mean $\mu$, producing more stable estimates. Clusters are then labeled \textit{right-leaning} if $\hat{\theta}_c \ge 0.5$, \textit{left-leaning} if $\hat{\theta}_c \le -0.5$, and \textit{neutral} if $-0.5 < \hat{\theta}_c < 0.5$. Clusters with no news citations are labeled \textit{no signal}.

\definecolor{darkred}{RGB}{140, 0, 0}
\definecolor{lightred}{RGB}{235, 110, 110}
\definecolor{darkblue}{RGB}{0, 0, 140}
\definecolor{lightblue}{RGB}{110, 110, 235}

\begin{table}[t]
\centering
\caption{Top 5 most cited news sources (and their bias scores) by platform. Bias scores are also indicated by color: \textcolor{darkblue}{far-left}, \textcolor{lightblue}{left}, \textcolor{lightred}{right}, and \textcolor{darkred}{far-right}.}
\begin{tabular}{@{}|@{\hspace{3pt}}l@{}r|l@{}r@{\hspace{3pt}}|@{}}
\hline
\multicolumn{2}{|c|}{\textbf{Bluesky}} & \multicolumn{2}{c|}{\textbf{Truth Social}} \\
\hline
\textbf{News Source} & \textbf{Count} & \textbf{News Source} & \textbf{Count} \\
\hline
\textcolor{darkblue}{RawStory (-2)} & 3.2K & \textcolor{lightred}{Rumble (+1)} & 10.1K \\
\textcolor{darkblue}{New York Times (-2)} & 3.2K & \textcolor{darkred}{Gateway Pundit (+2)} & 3.7K \\
\textcolor{darkblue}{The Guardian (-2)} & 2.4K & \textcolor{darkred}{Breitbart (+2)} & 2.1K \\
\textcolor{lightblue}{Washington Post (-1)} & 1.6K & \textcolor{darkred}{Fox News (+2)} & 1.6K \\
\textcolor{lightblue}{CNN (-1)} & 1.4K & \textcolor{darkred}{New York Post (+2)} & 1.1K \\
\hline
\end{tabular}
\label{table:news_counts}
\end{table}

\subsubsection{Social Dimension Construction.}
To construct a social dimension corresponding to political partisanship, we apply the framework of \citet{waller_quantifying_2021}, summarized in Figure \ref{fig:method}. First, seed cluster pairs are selected algorithmically by computing pairwise cosine similarities between the centroids of all left-leaning and right-leaning clusters. Pairs with high centroid similarity are presumed to differ primarily along the partisan dimension rather than in content, making them ideal seed pairs. Thus, we select the top 4 seed pairs based on cosine similarity. We chose four pairs after testing between one and nine seed pairs. Our analysis showed four pairs produced the most robust output.

For each seed pair of clusters, we compute the L2-normalized centroids of the two clusters as the mean of their post embeddings, where $\hat{\mathbf{r}}_i$ and $\hat{\mathbf{l}}_i$ denote the normalized centroids of the right-leaning and left-leaning clusters respectively. This forms the vector difference $\mathbf{d}_i = \hat{\mathbf{r}}_i - \hat{\mathbf{l}}_i$. A partisanship axis is then defined as the mean of these per-pair differences:
\begin{equation*}
    \mathbf{d} = \frac{1}{n} \sum_{i=1}^{n} \mathbf{d}_i
\end{equation*}
Each post $p$ is assigned a raw partisanship score $s_p$ by projecting its L2-normalized embedding $\hat{\mathbf{e}}$ onto $\mathbf{d}$:
\begin{equation*}
    s_p = \hat{\mathbf{e}} \cdot \mathbf{d}
\end{equation*}
Raw scores are standardized to z-scores across all posts, generating a final partisanship score $z_p$ for each post. Positive values indicate alignment with a right-leaning position; negative values indicate alignment with a left-leaning position.

To characterize the dominant topics of seed clusters, we drew five independent random samples of up to 300 posts and prompted \texttt{GPT-4o-mini}~\cite{hurst2024gpt} to generate a concise summary (2--3 sentences) and a descriptive label (3--5 words) for each sample. The summaries and labels were consistent across samples and manually verified by the authors. The summarization prompt is provided in the Appendix. 

Table \ref{tab:seed_pairs} reports the selected pairs together with their cluster characteristics and dominant topics. The pairs exhibit topic coherence: each pair addresses related political themes but frames them from opposing partisan perspectives. For example, the first seed pair, the left-leaning cluster is dominated by an intra-party critique---calls to action urging the Democratic party to reform its strategy---while its right-leaning counterpart adopts a more dismissive tone, ridiculing Democratic performance. Similarly, in the second pair, the left-leaning discourse centers on criticism of Biden's handling of the Israel-Palestine conflict, while the right-leaning cluster pivots to broader anti-immigration rhetoric.

\newcommand{\Lscore}[1]{\textcolor{darkblue}{#1}}
\newcommand{\Rscore}[1]{\textcolor{darkred}{#1}}

\begin{table}[t]
\centering
\caption{Characteristics of the chosen seed pair clusters.}
\label{tab:seed_pairs}
\begin{tabularx}{\columnwidth}{@{}crr>{\raggedright\arraybackslash}X@{}}
\hline
\textbf{Score} & \textbf{\#Cit.} & \textbf{Sim.} & \textbf{Topic} \\
\hline
\Lscore{$-0.87$} & 640  & \multirow{2}{*}{0.70} & Intra-party critique of Democratic political strategy \\
\Rscore{$+0.78$} & 958  &                        & Anti-Democrat rhetoric/hate \\
\hline\hline
\Lscore{$-0.69$} & 2687 & \multirow{2}{*}{0.56} & Pro-Palestine, Biden-Israel criticism \\
\Rscore{$+0.60$} & 1024 &                        & Immigration \& border security outrage \\
\hline\hline
\Lscore{$-1.13$} & 841  & \multirow{2}{*}{0.48} & Criticism of MAGA \\
\Rscore{$+0.96$} & 947  &                        & Conservative outrage and patriotism \\
\hline\hline
\Lscore{$-0.73$} & 1441 & \multirow{2}{*}{0.14} & Trump controversies and criticism \\
\Rscore{$+0.89$} & 3097 &                        & Conspiracy theories \& Extremism \\
\hline
\end{tabularx}
\vspace{2pt}

{
Each consecutive rowpair shares a \textit{Sim.} value, and pairs are ordered by descending cosine similarity.
\textcolor{lightblue}{Blue} and \textcolor{lightred}{red} \textit{Score} values indicate left- and right-leaning clusters respectively, based on mean media bias scores.
\textit{Sim.} is the cosine similarity between cluster centroids.
\textit{\#Cit.} is the count of news citations in the cluster.
}
\end{table}

\subsection{Validation}
We validate the partisanship axis in two ways, both of which exploit the fact that posts citing news sources carry an independent ground-truth signal---the AllSides bias score of the cited outlet---against which our projection-based partisanship scores can be compared.

\subsubsection{In-distribution Validation.}
Posts containing links to a news source that were not included in the Bluesky and Truth Social seed pair clusters served as a hold-out test set. For each such post, we computed its partisanship score by projecting its embedding onto the constructed partisanship axis and assessing whether its score is positively correlated with the media bias score of the cited news source. Because the axis was constructed using only the cluster centroids, and had no access to individual post-level citation scores, a strong correlation would indicate that the axis generalizes from the cluster level, where it was defined, to the post level, where it is applied.

\subsubsection{Out-of-distribution Validation.}
We further evaluate whether an axis generalizes beyond the Bluesky and Truth Social platforms using an independent dataset of posts drawn from X during the 2024 U.S. election period \citet{balasubramanian_public_2025}. The dataset was collected using a political keyword filter similar to ours. We applied the same preprocessing pipeline described in the Data section. Roughly 848K of the cleaned tweets contain a link to a news source rated by AllSides. We embed these posts using the same sentence encoder and project them into the same embedding space via the PCA fit on our Bluesky and Truth Social data. This produces embeddings that are comparable to those used to define the axis. We then score each post by projecting its embedding onto the partisanship axis and again test for a correlation with the media bias scores of the cited news sources. We are interested in the direction of the correlation: a positive correlation demonstrates that the partisan signal encoded in the axis survives transfer to an unseen platform. A stronger platform-specific correlation could, in principle, be recovered by re-fitting the axis on X data using the same procedure.

\subsubsection{Sensitivity to Seed Pair Selection.}
\begin{figure}[t]
    \centering
    \includegraphics[width=0.5\textwidth]{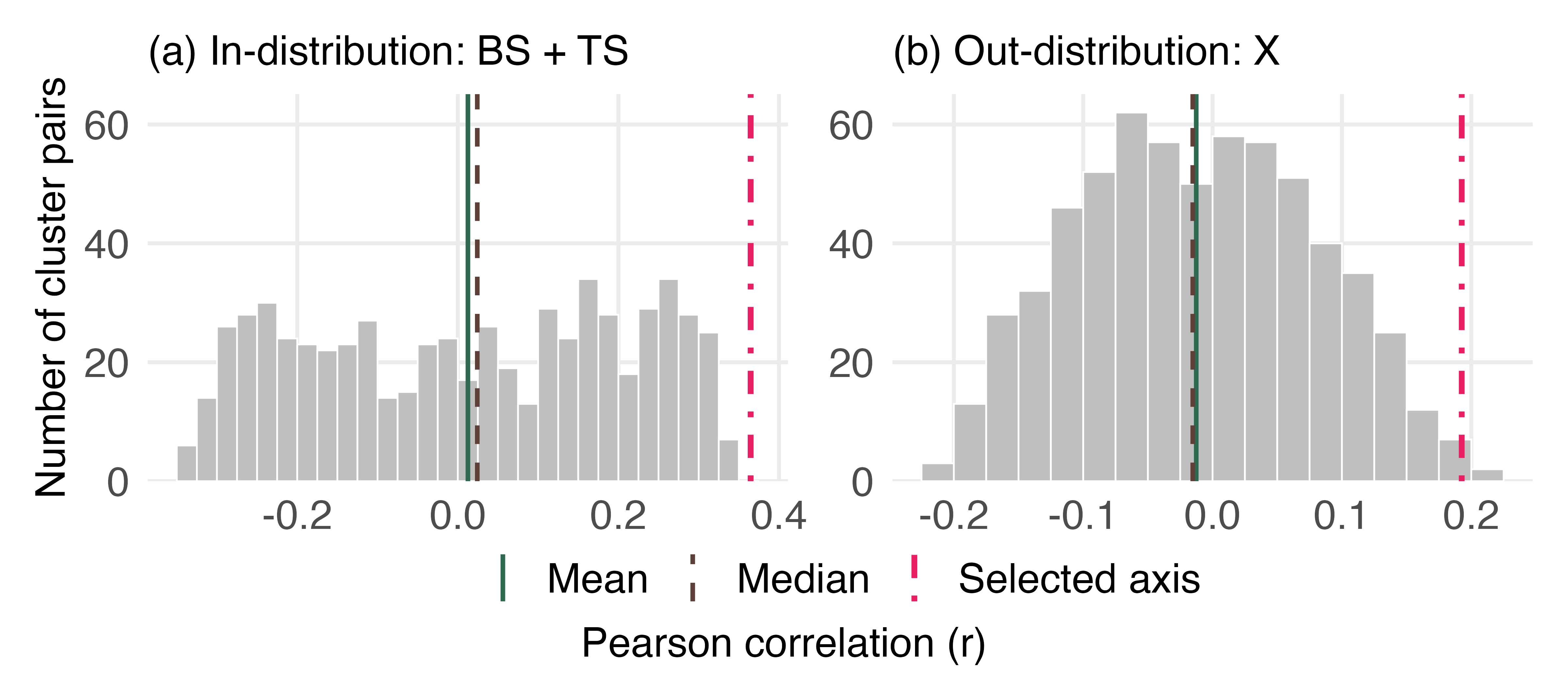}
    \caption{Sensitivity of the partisan axis to seed pair selection. Each bar represents the number of cluster pairs whose corresponding partisanship axis achieves a given Pearson correlation ($r$) with the media bias score, across all candidate axes. Panel (a) shows the in-distribution evaluation on held-out Bluesky (BS) and Truth Social (TS) posts; panel (b) shows the out-of-distribution evaluation on X.}
    \label{fig:seed_pair_sensitivity}
\end{figure}
The validations above are intended to demonstrate that our chosen seed pairs produce an axis where media bias scores correlate with our assigned partisanship scores. To verify that this correlation reflects the validity of our seed selection procedure rather than a property shared by arbitrary cluster pairings we conducted an exhaustive baseline experiment. For every possible pair of clusters $(c_i, c_j)$ in the corpus---not only those identified by the seed selection procedure, and not only pairs from opposing clusters---we constructed an axis from the difference of the L2-normalized centroids, projected all held-out news-citing posts onto the axis, and computed the Pearson correlation between the resulting partisanship scores and the media bias score of the cited news source. Posts belonging to either of the two clusters defining the axis are excluded to preserve the hold-out structure. The same procedure is then repeated on the X dataset.

This yields, for both evaluation sets, a full distribution of correlations across all $\binom{36}{2} = 630$ possible cluster pairings. If our seed selection procedure contributes meaningful information, the correlations for our selected pairs should fall in the upper tail of this distribution---the Results section will show that this is indeed the case.

\begin{figure*}[t]
    \centering
    \includegraphics[width=\textwidth]{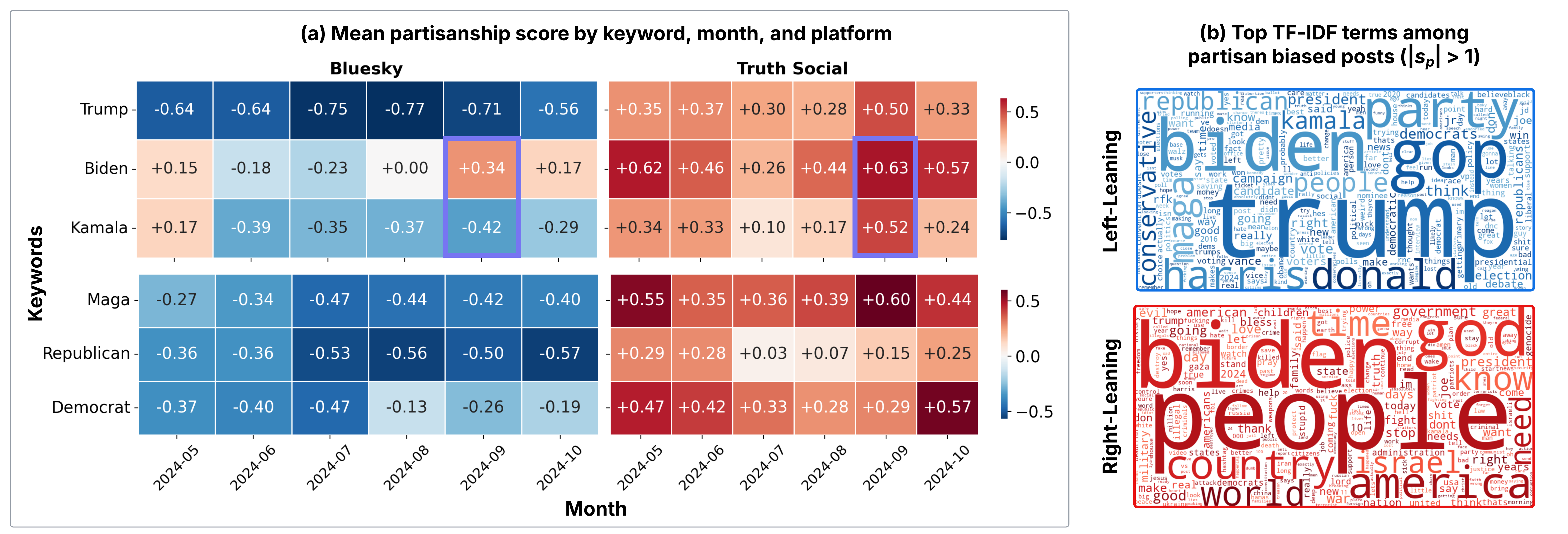}
    \caption{\textbf{(a)} Mean partisanship score ($z_p$) of posts containing each keyword, stratified by platform and month. Negative scores (blue) indicate left-leaning partisan alignment; positive scores (red) indicate right-leaning alignment. Keywords are grouped into candidate names (top) and party-affiliation terms (bottom). Highlighted with a purple border are the scores for the Democratic candidates in the month of September, when the first presidential debate happened. \textbf{(b)} TF--IDF wordclouds of the most discriminative terms between strongly left-leaning ($z_p < -1$) and strongly right-leaning ($z_p > 1$) posts across both platforms. Word size is proportional to TF--IDF weight.}
    \label{fig:heatmap_wordcloud}
\end{figure*}
\subsection{Qualitative Analysis}
To assess whether the resulting partisanship scores are coherent, we conduct two complementary qualitative analyses. First, we stratify posts by politically relevant keywords (e.g., \textit{Kamala}, \textit{Democrat} on the left; \textit{Trump}, \textit{Republican} on the right) and compare mean partisanship scores across platforms, by month. Keywords are identified by case-insensitive regex matching on the cleaned text. Prior to matching, different references to the same entity are collapsed into a single term (\textit{joebiden}, \textit{jbiden} $\rightarrow$ \textit{Biden}; the full mapping is provided in the Appendix). Mean scores are visualized as a heatmap in Figure~\ref{fig:heatmap_wordcloud}(a). Since the partisan axis is fit once and all posts are then scored using it, the monthly variation in the heatmap reflects how discourse moves along that fixed axis across time. Second, to characterize the language driving the partisan signal itself, we extract posts in the upper and lower tails of the score distribution ($|z_p| > 1$) and use term frequency-inverse document frequency (TF-IDF) to identify the terms that best discriminate between the two tails \cite{ramos2003using}. The resulting wordclouds are shown in Figure~\ref{fig:heatmap_wordcloud}(b).

\section{Results}

\subsection{Partisanship Dimension Validation}
The resulting partisanship scores are positively correlated with the media bias scores of cited news sources on both evaluation sets ($r = 0.365$, 95\% CI $[0.357, 0.373]$, $n = 43{,}919$, $p < 0.001$ for Bluesky and Truth Social; $r = 0.193$, 95\% CI $[0.191, 0.195]$, $n = 848{,}081$, $p < 0.001$ for X), indicating that the constructed partisan axis captures ideological bias across platforms. Both correlations are in the expected direction: posts citing right-leaning outlets score higher on the partisan axis, and posts citing left-leaning outlets score lower. The fact that the correlation remains positive and statistically significant on X---a platform the axis was never exposed to during fitting---supports the interpretation that the axis captures a general partisan signal rather than a platform-specific one. A stronger correlation could in principle be recovered by re-fitting the axis on X data using the same procedure.

\subsubsection{Sensitivity to Seed Pair Selection.}
Figure~\ref{fig:seed_pair_sensitivity} shows the distribution of the correlations achievable by axes constructed from all $630$ possible cluster pairs, together with the correlation achieved by the axis constructed from our selected seed pairs. In the in-distribution evaluation, our axis achieves a correlation that exceeds that of every alternative axis. On the out-of-distribution X evaluation, our axis falls within the $99.5$th percentile.

This result establishes that arbitrary cluster pairs do not, in general, yield axes that produce comparable scores, and our seed selection process recovers a near-optimal partisan signal from the available data. The fact that this remains true for a corpus unavailable during seed pair selection---the X sample---further supports the conclusion that the axis captures partisanship rather than properties idiosyncratic to Bluesky or Truth Social.

\subsection{Trends in Partisan Language}

\subsubsection{Partisanship-Keyword Associations Over Time.}
Figure~\ref{fig:heatmap_wordcloud}(a) reveals several patterns consistent with the platforms' expected political orientations, alongside others that reflect the dynamics of the 2024 election cycle. Posts mentioning \textit{Trump} carry strongly negative partisanship scores on Bluesky and consistently positive scores on Truth Social, as would be expected: each platform's user base discusses the candidate from the perspective of its predominant political leaning. The pattern for \textit{Biden}, however, is more revealing. On Bluesky, Biden-mentioning posts hover near zero or trend slightly right-leaning through the summer, departing sharply from the platform's otherwise left-leaning baseline. This is consistent with the erosion of enthusiasm for Biden's candidacy in the months preceding his July 21 withdrawal. The score reaches its most right-leaning value in September---after his withdrawal---suggesting that Bluesky discussions of Biden in that period were predominantly critical rather than left-aligned. Posts mentioning \textit{Kamala} show an inverse pattern on Bluesky: scores become increasingly left-leaning from July onward, peaking in September around the first presidential debate between Harris and Trump. This period is highlighted in the figure with a purple border. On Truth Social, an opposing opinion appears: Biden and Kamala posts both spike rightward in July, most likely reflecting Republican reaction to the candidate transition.

The bottom three rows capture general party affiliation rather than opinions on specific candidates. \textit{MAGA} and \textit{Republican} pattern appear as one would predict---strongly right-leaning on Truth Social and strongly left-leaning on Bluesky, the latter reflecting that Bluesky users discussing MAGA-related content typically do so critically. \textit{Democrat} shows a similar asymmetry.

\subsubsection{Partisan Language at Two Extremes.}
The wordclouds in Figure~\ref{fig:heatmap_wordcloud}(b) reveal an asymmetry in the language of strongly partisan posts. The left-leaning extremes reference both Democrats and Republicans prominently (\textit{biden}, \textit{harris}, \textit{trump}). The right-leaning extremes, by contrast, are dominated by civic and moral vocabulary (\textit{people}, \textit{god}, \textit{country}, \textit{america}, \textit{world}, \textit{need}), consistent with prior characterizations of right-leaning online discourse during the 2024 cycle as emphasizing nationalist and religious framings.

\section{Discussion \& Conclusion}
A critical finding of this work is that our partisanship scores recover within-platform trends in partisanship that platform identity alone cannot explain. On Bluesky, an otherwise left-leaning platform, posts mentioning Biden drift lose the support of the left through the summer of 2024---reflecting the well-documented erosion of enthusiasm for his candidacy---and even reach right-leaning values in September, after his withdrawal. Posts mentioning Harris on the same platform, become more strongly left-aligned from late July onward, peaking around the first presidential debate. These are not patterns that would be recovered using a method that simply learns ``Bluesky-style language" versus ``Truth Social-style language"; they reflect dynamic partisan sentiment toward specific political actors, that shifts in response to events in the political cycle. 

The validation results support this interpretation from a different perspective. On hold-out posts citing news sources from Bluesky and Truth Social, projected partisanship scores correlate with AllSides bias scores ($r = 0.365$). On an out-of-distribution X corpus unavailable during axis construction, the correlation ($r = 0.193$) continues to point in the expected direction. We interpret this result as evidence that the partisan signal captured by an axis can be applied to a new social media platform. Our exhaustive baseline analysis reinforces this conclusion: among the possible axes constructible from arbitrary cluster pairs, our seed-selected axis lands at the 100th percentile on the in-distribution evaluation and the 99.5th percentile using posts on X, indicating that the partisan signal the axis recovers is a direct consequence of the seed selection procedure.

Importantly, because the pipeline operates only on post text, the same procedure applies to any platform where users share rated news sources at non-trivial rates. The practical payoff is that cross-platform comparisons of partisanship dynamics---how different platforms responded to the same political event, how partisan attention shifts over time, and how the same content reads on ideologically opposing platforms---become directly manageable.

\paragraph{Limitations.}
There exist several limitations to our approach. First, the analysis is U.S.-centric: the media bias scores our method relies on are provided by AllSides, which rates only U.S. news outlets. The constructed partisanship axis assumes a U.S. two-party context. Cross-national application to multi-party systems would require an analogous bias-rating tool. Second, the method is sensitive to two methodological choices: the clustering procedure, and the seed-pair selection that defines an axis. While our sensitivity analysis shows that the selected axis outperforms the majority of arbitrary pairings, the construction is not unique---more principled clustering procedures such as HDBSCAN might yield more semantically coherent clusters and better seed selection. Third, our chosen platforms (Bluesky and Truth Social) are politically asymmetric by design---a right-leaning and a left-leaning alt-tech platform---which makes them ideal for demonstrating portability across opposing environments. This leaves open how the method behaves on more ideologically heterogeneous mainstream platforms such as Reddit or Facebook. Beyond these limitations, downstream applications of the method---to questions of temporal polarization dynamics, cross-platform information flow, or platform-motivated radicalization---become possible once partisanship can be measured and compared across platforms. This represents the broader impact of our methodological contribution.


\bigskip

\bibliography{aaai2026}

\appendix

\section{Keywords Used for Data Collection}

Enumerated below is the full list of keywords used to collect posts on Bluesky.

\vspace{2pt}
\noindent\rule{\columnwidth}{0.4pt}
\noindent\small\textit{2024 Elections, 2024Elections, 2024 Presidential Election, 2024PresidentialElections, 2024USElections, US Elections, USElections, Biden, Joe Biden, JoeBiden, Joseph Biden, JosephBiden, Biden2024, bidenharris2024, Donald Trump, DonaldTrump, Trump2024, trumpsupporters, trumptrain, conservative, republicansoftiktok, democratsoftiktok, CPAC, GOP, KAG, MAGA, ultramaga, Nikki Haley, NikkiHaley, Ron DeSantis, RonDeSantis, RNC, DNC, thedemocrats, the democrats, Democratic party, Democraticparty, Republican party, Republicanparty, Third Party, ThirdParty, Green Party, GreenParty, Independent Party, IndependentParty, No Labels, NoLabels, Kamala Harris, KamalaHarris, Joe Biden and Kamala Harris, Marianne Williamson, MarianneWilliamson, Dean Phillips, DeanPhillips, williamson2024, phillips2024, RFK Jr, RFKJr, Robert F. Kennedy Jr., Jill Stein, JillStein, Cornel West, CornellWest, voteblue2024, letsgobrandon, makeamericagreatagain, Vivek Ramaswamy, VivekRamaswamy, Snowballing}
\vspace{2pt}
\noindent\rule{\columnwidth}{0.4pt}

\section{LLM Prompt Used for Cluster Summarization}

The following prompt was used to generate summaries and labels for each seed cluster via \texttt{GPT-4o-mini}. The placeholders \texttt{\{n\}} and \texttt{\{posts\}} are replaced with the sample size and sampled post text respectively.

\vspace{2pt}
\noindent\rule{\columnwidth}{0.4pt}
\begin{verbatim}
Below are {n} social media posts from the 
same cluster. Summarize what this cluster 
is about in 2-3 sentences, then assign it 
a short label (3-5 words).

Posts:
-----------------------------
{posts}
-----------------------------

Respond in this exact format:
LABEL: <label>
SUMMARY: <summary>
\end{verbatim}
\noindent\rule{\columnwidth}{0.4pt}

\section{Keyword Mapping for Entity References}
The following groups define how different references to the same entity are collapsed to one, prior to keyword matching.

\vspace{2pt}
\noindent\rule{\columnwidth}{0.4pt}
\begin{verbatim}
Trump      <- trump, donaldtrump, donaldjtrump,
              donjtrump, realdonaldtrump
Biden      <- biden, joebiden, josephbiden
Kamala     <- kamala, harris

Maga       <- maga, makeamericagreatagain, 
              ultramaga
Republican <- republican, gop, rnc, 
              conservative
Democrat   <- democratic, dnc, thedemocrats, 
              liberal
\end{verbatim}
\noindent\rule{\columnwidth}{0.4pt}

\end{document}